\documentclass{pasj00}
\draft

\RelativeFigurePath{figure}

\begin{document}
\SetRunningHead{Author(s) in page-head}{Running Head}
\Received{}%{yyyy/mm/dd}
\Accepted{}%{yyyy/mm/dd}

\title{Effects of Compton scattering on the Gamma Ray Spectra of Solar flares} 

%%% begin:list of authors
\author{Jun'ichi \textsc{Kotoku}} 
\affil{National Astronomical Observatory of Japan,
2-21-1 Osawa, Mitaka, Tokyo 181-8588, JAPAN}
\email{junichi.kotoku@nao.ac.jp}
\author{Kazuo \textsc{Makishima}\altaffilmark{1} and Yukari \textsc{Matsumoto}\altaffilmark{2}}
\affil{Department of Physics, University of Tokyo, 
Bunkyo-ku, Tokyo, 113-0022}
\and
\author{Mitsuhiro \textsc{Kohama}, Yukikatsu \textsc{Terada} and Toru \textsc{Tamagawa}}
\affil{RIKEN (Institute of Physical and Chemical research), Wako-shi, Saitama}
%%% end:list of authors

\altaffiltext{1}{Also at RIKEN}
\altaffiltext{2}{Present address: Mitsubishi Electric Co., Ltd.}

\KeyWords{solar flare: gamma-ray: Compton scattering}

\maketitle

\begin{abstract}
Using fully relativistic GEANT4 simulation tool kit, the transport
of energetic electrons generated in solar flares was Monte-Carlo
simulated, and resultant bremsstrahlung gamma-ray spectra
were calculated.  The solar atmosphere was approximated
by 10 vertically-stacked zones. The simulation took into account
two important physical processes,that the bremsstrahlung photons
emitted by precipitating relativistic electrons are strongly
forward beamed toward the photosphere, and that the majority of
these gamma-rays must be Compton back-scattered by the solar
atmosphere in order to reach the observer. Then, the Compton
degradation was found to make the observable gamma-ray spectra
much softer than is predicted by simple analytic calculations.
The gamma-ray signals were found to be enhanced by several
conditions, including a broad pitch-angle distribution of the
electrons, a near-limb flare longitude, and a significant tilt
in the magnetic field lines if the flare longitude is rather small.
These results successfully explain several important flare properties
observed in the hard X-ray to gamma-ray range, including in
particular those obtained with Yohkoh. 
A comparison of  the Yohkoh spectrum
from a GOES X3.7 class limb flare on 1998  November 22,
with a simulation assuming a broad electron pitch-angle distribution, 
suggests that gamma-rays from this particular solar flare were 
a mixture  of direct bremsstrahlung photons
and their Comptonization. 
\end{abstract}

%%%%%%%%%%%%%%%%%%%%%%%%
\section{Introduction}\label{sec.introduction}
%%%%%%%%%%%%%%%%%%%%%%%%

Solar flares are sudden, brief, and powerful outbursts, 
releasing an energy of $10^{28-34}$ ergs on time scales 
of a second to several tens of minutes. 
In solar flares, both protons and electrons are accelerated 
to non-thermal energies. 
Although these energetic charged particles are strongly affected by the 
solar magnetic fields and cannot usually be detected, 
we can utilize neutral secondary particles, neutrons and photons, 
to probe them.
Among them, hard X-ray to gamma-ray continua, produced 
via bremsstrahlung process, provide the best diagnostics of the 
flare-accelerated electrons.  

Although high-spatial-resolution images of flares in the hard X-ray and 
gamma-ray energies have become available by the Yohkoh HXT \citep{Kosugi1991}
and RHESSI \citep{RHESSI2004} respectively, 
we still depend heavily on the spectral data when attempting to study 
the non-thermal electrons. 
Actually, to calculate back the electron spectrum from the observed 
gamma-ray spectra  has been an important issue in solar physics.  
As we understand, flare-accelerated electrons having a power-law 
distribution penetrate the solar 
chromosphere, and produce bremsstrahlung photons via ``thick-target'' emission. 
The thick-target emission is a process wherein non-thermal electrons 
impinge on a thick matter, and lose almost all their energies 
through repeated Coulomb collisions while emitting X-ray/gamma-ray photons 
via bremsstrahlung. 

In the thick-target condition, 
the spectrum of X-ray emitting electrons is formed by 
an equilibrium between the injection of newly accelerated electrons 
and the loss of their energies predominantly through collisions. 
If the injected electron number spectrum is given by a power-law as 
$F(E)\sim E^{\,-\delta}$ (electrons keV$^{-1}$ s$^{-1}$), 
with $E$ the electron kinetic 
energy and $\delta$ a constant called electron index, 
then we expect the X-ray photon-number spectrum also to take a power-law 
from as 
\begin{equation}
\frac{dJ(h\nu)}{d(h\nu)} \sim (h\nu)^{-\gamma}, 
\label{eq.thickTarget}
\end{equation}
where $\gamma$ is a quantity called photon index. 
In 1974, Brown found an approximate relation as 
\begin{equation}
\gamma = \delta - 1
\label{eq.Brown}
\end{equation}
in non-relativistic regime, assuming that all emitted photons are collected.  
The photon spectrum thus becomes relatively flat, 
because of ``loss-flattening'' effect working on the electrons; 
as they penetrate deeper into the target material, 
the spectrum gradually hardens because lower-energy electrons have shorter 
time scales of energy loss through Coulomb collisions.

Besides the thick-target condition, another extreme case called thin-target 
condition (Brown 1971) was studied extensively. 
This represents a situation 
wherein a bunch of energetic 
electrons pass through a thin medium while emitting bremsstrahlung photons,  
but they leave the emission region with their energies almost unchanged. 
In the thin-target condition, 
the energy distribution of the incident electron 
is not significantly affected by collisions with the target. 
The emergent photon spectrum is steeper, 
because the electrons are free from the loss flattening effect  
and is characterized as \citep{Brown1971} 
\begin{equation}
\gamma = \delta + 1/2. 
\label{eq.BrownThin}
\end{equation}

Since then, there have been many attempts to improve equations 
(\ref{eq.Brown}) and (\ref{eq.BrownThin}),
 particularly toward relativistic regime.   
Some utilized purely analytic calculations of various probability 
distributions, while others employed numerical solutions to the 
Fokker-Planck equation, or Monte Carlo techniques. 
In electron energies up to 10 MeV, \citet{Bai1982} investigated 
 angular dependence of the bremsstrahlung in the hard X-ray range,  
and discussed the associated electron transport. 
\citet{Murphy1987} dealt with the particle transport by employing 
a simple thick target emission model, 
assuming an isotropic electron distribution. 
\citet{Dermer1986} conducted a detailed study of the directional bremsstrahlung 
emission by anisotropically accelerated electrons. 
\citet{Miller1989} made a calculation including pitch angle scattering 
of electrons by plasma waves. 
These previous works have indicated that the gamma-ray continua of 
solar flares can be generally explained in terms of chromospheric 
thick-target emission, produced by electrons 
having a power-law energy distribution 
with $2 < \delta <4$. 

In spite of these extensive studies,
the estimation of electron spectra from flare
gamma-ray data would not be self-contained
unless we take into account another important physical process;
namely, Compton scattering.
The relativistic bremsstrahlung from precipitating electrons must be 
significantly forward-beamed, so that a majority of gamma-ray photons 
must be Compton back-scattered by the solar materials in order to reach us. 
This process will strongly degrade the gamma-ray energies, and affect the 
photon spectrum to be observed. 
To treat this kind of physical condition which involve 
multiple Compton scatterings, 
we clearly need to use Monte Carlo techniques. 

The effect of Compton scattering has been studied by several authors,
beginning, e.g., with \citet{BaiRamaty1978} 
who used Monte-Carlo simulations in such early days.
Kontar et al. (2006) considered this issue using
a semi-analytic Green's function by Magdzias \& Zdziarski (1995),
and calculated ``primary'' photon spectra from observed ones.
These works have clarified
that the Compton back-scattering can significantly
modify the bremsstrahlung spectra.
However, these pioneering works have treated electrons
with sub-MeV kinetic energies, 
leaving higher energies less studied.

In the present paper, we calculate not only the electron 
transport and bremsstrahlung emission, but also the photon propagation 
via Compton process.  
For this purpose, we employ a Monte Carlo simulation toolkit 
named {\sc Geant4}, which is widely used in experimental high energy physics. 
Our results imply
that the Compton scattering has significant effects 
on the observed gamma-ray spectra.
Together with full-relativistic effects,
this is expected to require revision to the relation 
such as equation(\ref{eq.Brown}) and equation(\ref{eq.BrownThin}). 
We show the relevance of these results to actual flares,
by comparing the simulated gamma-ray spectra  with 
those observed with Yohkoh from a sample of solar flares.

\clearpage
%%%%%%%%%%%%%%%%%%%%%%%%%%%%%%
\section{Method of Simulation}
%%%%%%%%%%%%%%%%%%%%%%%%%%%%%%

%--------------------------------------------------
\subsection{The {\sc Geant4} toolkit}
%--------------------------------------------------
The {\sc Geant4} Monte Carlo simulation toolkit \citep{Geant4}, 
developed in experimental high energy physicists, 
has four important features. 
Firstly, it allows a very flexible construction 
of ``geometry'' in which particles interact. 
Secondly, it can ``track'' individual particles while monitoring their physical 
quantities such as the energy, momentum, and position;   
the name {\sc Geant} in fact comes from ``GEometry ANd Tracking''. 
Thirdly, it allows us to incorporate the desired interactions;  
in the present case, Coulomb scattering and bremsstrahlung 
for electrons, and Compton scattering for photons. 
Finally, the toolkit has been tested and calibrated extensively by many investigators 
under different conditions.

%----------------------------------------------------------
\subsection{Particle injection, interaction, and tracking}
%----------------------------------------------------------

In our Monte Carlo simulation using {\sc Geant4}, 
we basically shoot a parallel beam of a large number of electrons, 
either vertically or slantly, into our ``solar atmosphere'' 
to be explained later. 
The electrons are assumed either to be mono-energetic, 
or to have a power-law energy distribution with an index $\delta$. 
In the latter case, the initial electron energies are assumed 
to be in the range of 1--100 MeV, so as to simulate gamma-ray flares. 

We implemented the following physics processes which we require our particles to obey in 
our simulation: 
ionization, multiple scattering, and bremsstrahlung for electrons,  
while photo-absorption, 
Compton scattering and electron-positron pair creation for photons. 
We suppressed the production of secondary electrons for simplicity. 
Because those secondary electrons (typically less than 100 keV) 
have much lower energies than the primary ones, 
they can be considered negligible.

One of the essential features of {\sc Geant4} is step-by-step tracking 
of each individual particle, considering all the implemented elementary processes. 
The particle tracking algorithm roughly consists of the following 4 steps. 
1) The initial particle velocity is calculated.  
2) Each physical process calculates the mean-free-path of the specified 
particle under the relevant interaction, and generates a random number 
around the mean. 
Then it proposes this random number as a step length. 
3) Among various competing processes, the interaction which proposes 
the shortest physical length is adopted. 
4) Track properties, such as the kinetic energy and momentum, 
of the current particle are updated. 

Among the implemented processes, the Coulomb scattering
of electrons need special consideration, because its mean
free path is much smaller than those of the other processes
(e.g., $\sim 10^{-2}$ of that of bremsstrahlung for a 10 MeV
electron). As a result, its full Monte Carlo treatment would
demand too much computing times.  
Therefore, the expected total energy loss of
an electron due to multiple Coulomb scatterings,
in each step which is determined by the processes
other than the Coulomb scattering itself,
is computed based on semi-empirical formulae by \citet{Messel1970}.
Fluctuations around the mean energy loss is expressed
by generating random numbers after \citet{Kati1995}.
Similarly, the total spatial displacement caused by
the multiple scattering in each step is calculated
based on Lewis method \citep{Lewis1950},
which solves a diffusion equation of electrons in the matter.
Fluctuations around the expected mean is again
represented by random numbers \citep{Geant4}.
Thus, {\sc Geant4} simulates the effect of Coulomb interactions
after a given step as a statistical sum of a large number
of scatterings, instead of treating them one by one.

In this way, the injected electrons and the bremsstrahlung-generated photons 
are tracked, until they escape out of the model boundary, 
or they become less energetic than a certain threshold energy 
which we set at 1 keV. 
We collect these photons under a specified condition, and make their spectra. 

\begin{figure}[htbp]
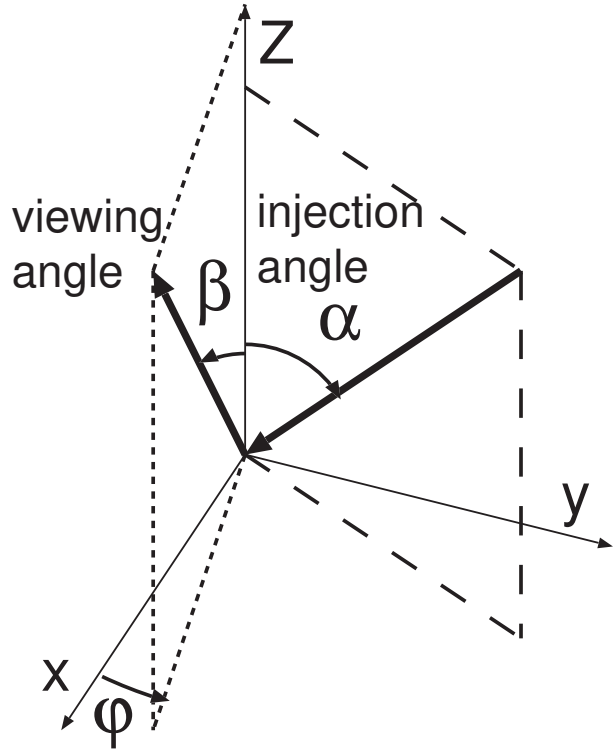

\begin{center}
\FigureFile(8cm,10cm){coodinate.eps}
\caption{A coordinate system in the present simulation. 
The normal to the solar photosphere is $z$ axis, 
and the electron injection angle measured from $+z$ direction 
is denoted $\alpha$. 
The viewing direction is specified by a polar angle $\beta$ 
and an azimuth angle $\varphi$. 
}
\label{fig.coordinate}
\end{center}
\end{figure}

The simulation of flare-accelerated electrons, 
precipitating onto the chromosphere, 
requires yet another physical process to be implemented: 
electron gyration around the solar magnetic fields. 
Its exact treatment, however, would make the simulation extremely 
time-consuming, because the gyration radius of a $\sim 100$ MeV electron 
in a $\sim 100$ Gauss field is as small as $\sim 10$ m. 
Therefore, we have decided not to take gyration into our simulation, 
but instead, to consider slant injections; 
as illustrated in figure \ref{fig.coordinate}, 
injecting electrons at a polar angle $\alpha$, 
and collecting the emergent photons at another polar angle $\beta$ 
but all over the azimuth angle $\varphi$, are equivalent 
to simulating electrons gyrating with a constant pitch angle $\alpha$,  
as long as the field lines are close to normal to the photosphere. 
The case of tilted magnetic fields can be reproduced by a superposition of 
different injection angles, $\alpha$.

%--------------------------------------------------
\subsection{Electron transport under a simple geometry}
%--------------------------------------------------

Before actually simulating the electron transport and gamma-ray emission 
employing a realistic model for the solar atmosphere, 
we carried out simpler simulations with two purposes in mind; 
to validate our simulation code by examining whether each of the 
relevant physical processes is correctly reproduced, 
and to grasp the essence of the physics to be investigated. 
We accordingly approximated the sun as a hydrogen gas box of 
$10,000\ \mathrm{km} \times 20,000\ \mathrm{km} \times 20,000\ \mathrm{km}$,  
having a uniform density of 3.2$\times 10^{-7}$ g cm$^{-2}$  
which is the density at the solar photosphere. 
Hereafter we call this model ``simple solar atmosphere model''; 
it is one of three uniform target models employed in the present paper. 

\begin{figure}[htbp]
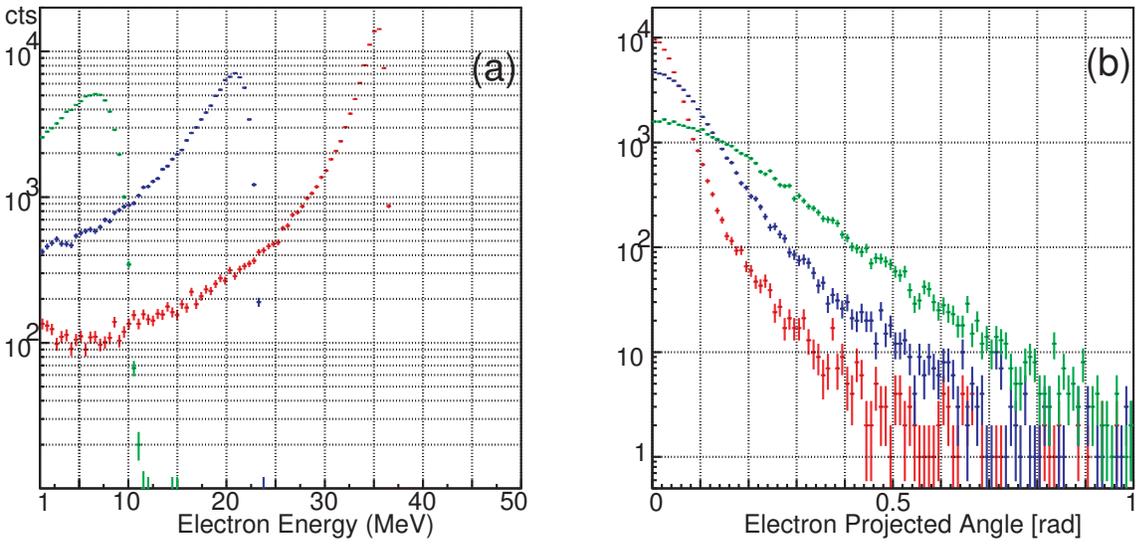

\begin{center}
\FigureFile(8cm,8cm){v9.93_50MeVElectronSpectrum.eps}
\FigureFile(8cm,8cm){v9.93_50MeVElectronProjAngDist.eps}
\caption{A {\sc Geant4} simulation of electrons with an initial kinetic energy 
of 50 MeV, injected vertically into a hydrogen target with a uniform density 
of $3.2\times10^{-7}$ g cm$^{-3}$ (the simple solar atmosphere). 
Data are collected at depths of 100 km (red), 200 km (blue), 
and 300 km (green). 
(a) The energy spectrum. 
(b) The distribution (per unit projected angle) 
of angular deflection, 
measured from the initial direction of injection.}
\label{fig.monoElectronSpectrum}
\end{center}
\end{figure}

We vertically injected $10^5$ mono-energetic electrons with an initial kinetic energy 
of 50 MeV into our ``simple solar atmosphere''. 
In order to monitor how the Coulomb interaction affects the spectrum and 
angular distribution of the electrons, 
we prepared imaginary boundaries at different depths, 100, 200, and 300 km 
from the injection surface,  
and collected electrons at each boundary.
Figure \ref{fig.monoElectronSpectrum} 
a shows the electron spectra obtained in this way, 
while Figure \ref{fig.monoElectronSpectrum}b the associated angular distributions. 
From these, we can see that the maximum energy of the spectrum 
decreases as the electrons penetrate deeper. 
Since a 100 km thick slab in our ``solar atmosphere'' has a column density of 
3.2 g cm$^{-2}$, an electron of energy $\sim$ 50 MeV passing through 
it is expected to lose $\sim$ 15 MeV,  
as we can easily calculate from the Bethe-Bloch formula.  
The results in Figure \ref{fig.monoElectronSpectrum}a indeed meet this expectation. 
The geometrical thickness (10,000 km) of the model is thus 
large enough to make it physically thick to Coulomb loss, 
for electrons with initial energies up to $\sim$1.5 GeV. 
Furthermore, the electron energy distribution is seen to broaden 
as the energy loss proceeds. 

Figure \ref{fig.monoElectronSpectrum}b reveals that the injected electrons, 
initially  forming a parallel beam, 
are gradually deflected due to Coulomb interactions with ambient atoms. 
The simulated angular distributions approximately consist of two components: 
a Gaussian component caused by small-angle multiple scatterings,  
and a power-law tail due to large-angle Rutherford scatterings. 

In a similar way, we simulated the transport of energetic electrons  
which has a power-law distribution over the 1-100 MeV range with $\delta=1.2$. 
The spectra and angular distributions, measured at depths of 100, 200, 300, 
400, and 500 km, are presented in figure \ref{fig.powElectronSpectrum}. 
The maximum electron energy (initially at 100 MeV) thus decreases 
according to the Bethe-Bloch formula. 
Furthermore, the low-energy part of the spectrum gradually flattens, 
because the energy loss per unit length is roughly independent of the 
electron energy, and hence lower-energy electrons suffer a larger 
{\it fractional} energy loss than higher-energy ones. 
This is essentially the same as the ``loss flattening'' effect 
mentioned in section \ref{sec.introduction}. 

The electron index we employed here, $\delta=1.2$, 
is significantly smaller (flatter) than would be achieved 
in the standard  diffusive shock acceleration process 
\citep{Blandford1978,Bell1978a}. 
This is motivated by two reasons. 
One is that the effects of Compton scattering are more clearly observed 
as $\delta$ gets smaller. 
The other is that such a flat electron distribution could be realized 
in actual solar flares, via, e.g., direct electric acceleration. 
Hereafter, we hence consider both flat ($\delta<2$) and steep ($\delta>2$) 
electron distributions. 

%----------------------------------
\subsection{Bremsstrahlung spectra}
%----------------------------------

We investigated photon spectra created by the bremsstrahlung process 
under the two representative conditions,  
namely the thick-target and thin-target conditions explained in 
section \ref{sec.introduction}.
The {\sc Geant4}-simulated gamma-ray spectra are compared with 
analytic formulae, which are fully relativistic unlike the non-relativistic 
calculations of eq.(\ref{eq.Brown}) and eq.(\ref{eq.BrownThin}).

\subsubsection{Thin-target emission}
To validate the bremsstrahlung process under the thin-target condition 
which is simpler of the two,  
we let $10^8$ mono-energetic electrons vertically penetrate 
a second simplified model atmosphere, namely a uniform 
hydrogen gas with a size of (1000 km)$^3$, 
having a density of $1.0\times 10^{-11}$ g cm$^{-3}$; 
this is henceforth called ``thin atmosphere model''. 
The model has a column density of 1.0 mg cm$^{-2}$,  
so that the injected electrons lose their energies 
by no larger than 1 \%, in agreement with the thin-target 
condition. We collected the information of each bremsstrahlung photon  
when it was created, and obtained the spectra shown  
in  figure \ref{fig.BremsMonoEnergySpectrum}. 

Also shown in  figure \ref{fig.BremsMonoEnergySpectrum} are analytically 
calculated spectra, 
using a fully relativistic formula  
\citep{Schiff1951,KochandMotz1959}, 
as  
\begin{eqnarray}
d\sigma_k &= \frac{2 Z^2 r_e^2}{137}\frac{dk}{k}
        \left\{\left(1+\left(\frac{\epsilon}{\epsilon_0}\right)^2 - \frac{2\epsilon}{3\epsilon_0}\right)
        \left(\ln M(0) + 1 - \frac{2}{b}\tan^{-1}b\right)\right. \nonumber\\
&\quad+\, \left. \frac{\epsilon}{\epsilon_0}\left[
        \frac{2}{b^2}\ln(1+b^2) +
        \frac{4(2-b^2)}{3b^3}\tan^{-1}b - \frac{8}{3b^2} +
        \frac{2}{9}\right] \right\}
\label{KM3BSe}
\end{eqnarray}
with
\begin{equation}
\qquad b = \frac{2 \epsilon_0 \epsilon Z^{1/3}}{111k}, \qquad
\frac{1}{M(0)} = \left(\frac{k}{2 \epsilon_0 \epsilon}\right)^2 + \left(\frac{Z^{-1/3}}{111}\right)^2.
\end{equation}
Here, $d\sigma_k/dk$ is a cross section for an electron of total energy 
$\epsilon_0 m_e c^2$ (including the rest mass energy) to interact with 
a target atom of charge $Z$, and emit a bremsstrahlung photon of energy 
$k m_e c^2$, thus achieving a final energy of 
$\epsilon m_e c^2 \equiv (\epsilon_0 - k) m_e c^2$. 
The classical electron radius is denoted $r_e$. 

As seen in figure \ref{fig.BremsMonoEnergySpectrum}, 
the difference between our simulation and 
the analytical calculation is less than 10 \%, 
indicating that the bremsstrahlung process has been correctly implemented. 
In the fully relativistic regime($\epsilon_0 \gg 1$,$\epsilon \gg 1$), 
in particular, 
the emitted gamma-ray spectrum has a power-law shape with a photon index 
$\gamma \simeq 1.0$, or an energy index of $\simeq 0$, 
up to the maximum photon energy that is identical to 
the initial photon energy. 
This is reasonable, because eq.(\ref{KM3BSe}) predicts
$d\sigma_k/dk \propto k^{-1}$ when $\epsilon_0 \sim \epsilon$ 
and hence $b \gg 1$.

%PL thin
Figure \ref{fig.powerThinSpectrum} shows the bremsstrahlung photon spectra 
calculated under the same thin-target condition, 
but when the electrons are assumed to have a power-law spectrum distributed from 1 to 100 MeV, with $\delta=0.8$ or 1.5. 
The simulated spectra again exhibit approximately power-law like shapes,   
with a steeper slope than in the case of mono-energetic electrons 
because softer electrons can obviously emit only lower-energy photons.   

In figure \ref{fig.powerThinSpectrum}, 
we also show photon spectra calculated analytically using equation (\ref{KM3BSe}) as
\begin{equation}
N(k) = n_i \int^{E_{\mathrm{max}}}_{k+1} v(\epsilon_0) J_e(\epsilon_0) \frac{d\sigma}{dk} d \epsilon_0 ,
\label{eq.PowBremsThin}
\end{equation}
where $N(k)$ is the photon number spectrum, 
$J_e(\epsilon_0)$ is the injected electron spectrum 
($\propto \epsilon_0^{-\delta}$ in the present case),   
$n_i$ is the target ion density, and 
$v(\epsilon_0)$ is the velocity of an electron having an energy $\epsilon_0$. 
The simulation agrees with the analytic results within $\sim 20\%$. 
Thus, the {\sc Geant4} simulation and the fully-relativistic analytic 
formalism consistently indicate $\gamma \simeq 1.3$ for $\delta = 0.8$ and 
$\gamma \simeq 1.6$ for $\delta = 1.5$, both measured in a typical gamma-ray 
energy range of 1 to 10 MeV. 
These photon indicies are significantly smaller (flatter),  
particularly for $\delta = 1.5$, when compared to the conventional 
non-relativistic relation of equation (\ref{eq.BrownThin}). 

In this way, we repeated the {\sc Geant4} simulation with difference
values of the electron index,
and calculated the Bremsstrahlung photon index
over an energy of 1-10 MeV.
The obtained relation between
the electron index and the photon index is shown
in Figure \ref{fig.thinthick},
in comparison with the prediction by equation (\ref{eq.BrownThin}).
We see that the formula of equation (\ref{eq.BrownThin})
in the non-relativistic regime,
namely $\gamma = \delta + 1/2$, is no longer valid in the present relativistic
regime, with the discrepancy increasing toward larger electron indicies.

The cross section of bremsstrahlung as a function of emitted photon energy
and angle is given by \citet{Schiff1951} and \citet{KochandMotz1959} as  
                                                                                
\begin{equation}
d\sigma_{k,\theta_0} = \frac{4 Z^2 r_e^2}{137}\frac{dk}{k}ydy
        \left\{\frac{16y^2 \epsilon}{(y^2+1)^4\epsilon_0} - \frac{(\epsilon_0+\epsilon)^2}{(y^2+1)^2\epsilon_0^2}
        + \left[\frac{\epsilon_0^2 + \epsilon^2}{(y^2+1)^2 \epsilon_0^2}
        - \frac{4y^2 \epsilon}{(y^2+1)^4 \epsilon_0}\right]\ln M(y)\right\},
\label{KM2BS}
\end{equation}
where ``reduced photon angle'' $y$, together with its function $M(y)$, is given by
\begin{equation}
y = \epsilon_0\theta_0,\
\frac{1}{M(y)} = \left(\frac{k}{2 \epsilon_0 \epsilon}\right)^2
        + \left(\frac{Z^{1/3}}{111(y^2+1)}\right)^2
\end{equation}
and $\theta_0$ is the initial angle of an emitted photon. 
Figure \ref{fig.BremsAngDistValid} shows equation (\ref{KM2BS}) 
as a function of $\epsilon_0\theta_0$,  
where the results of the simulation using the thin atmosphere model are superposed. 
Thus, the whole emission becomes concentrated roughly
within $\epsilon_0 \theta_0 \leq 1$;
the emission is strongly forward-peaked to within
$|\theta_0|\leq \epsilon_0^{-1}$.

\subsubsection{Thick-target emission}
The thick-target emission is significantly more complex 
than the thin-target case,  
because it results from electrons while their energies change 
continuously due to Coulomb scattering. 
Moreover, it has a higher practical importance, 
because the flare hard X-rays and gamma-rays are thought to be produced 
mainly through this process 
in the solar chromospheric and denser regions.  

Under the thick-target condition, we first investigated the emission 
from mono-energetic electrons, 
as we did for the thin-target case.  
We shot $10^7$ electrons with an initial kinetic energy of 10 MeV 
or 50 MeV, into a hydrogen target of which the size is (10$^4$ km)$^3$; 
this is our third simplified model atmosphere called ``thick atmosphere model''. 
Figure \ref{fig.MonoThickSim} shows the spectra 
produced by collecting all photons emitted by each electron 
while its energy gradually decreases down to $\lesssim 1$ keV. 
Compared with the thin-target emission from mono-energetic electrons, 
the photon spectra are somewhat steeper, and roll over more prominently 
toward the maximum energy. 
Obviously, this is because the electrons keep losing their energies 
while radiating. 

In figure \ref{fig.MonoThickSim}, 
we also show the results of analytic calculations, 
which are formulated as 
\begin{eqnarray}
N(\epsilon_\mathrm{ini},k) &= n_i \int^{t|_{\epsilon = k+1}}_{t|_{\epsilon = \epsilon_\mathrm{ini}}}
v \frac{d\sigma}{dk}\, dt \nonumber \\
&= n_i \int^{\epsilon_{\mathrm{ini}}}_{k+1} v
\left(\frac{d\epsilon}{dt}\right)^{-1} \frac{d\sigma}{dk}\, d\epsilon ,
\label{eq.MonoBremsThick}
\end{eqnarray}
where $t$ is the time, 
$\epsilon_\mathrm{ini}$ is the initial total energy of an electron, 
$d\epsilon/dt$ is its energy loss rate,  
$v$ is the instantaneous electron velocity, 
$n_i$ is the ion density of the target plasma, 
and $\frac{d\sigma}{dk}$ refers to equation (\ref{fig.BremsMonoEnergySpectrum}). 
Again, the simulation and analytic calculation agrees within $\le 20$\%. 

%PL thick
We investigated the thick-target emission from power-law distributed electrons, 
with $\delta=0.8$ or $1.5$, over the 1-100 MeV energy range. 
Employing $10^6$ electrons,  
we collected information of the photons in the same way 
as in the preceding simulations, and obtained figure \ref{fig.PowThickSim}. 
Designating the electron injection rate (electrons $\mathrm{cm}^{-2} \mathrm{s}^{-1}$)
by $f(\epsilon_\mathrm{ini})$ and the target area by $A$, 
and employing equation (\ref{fig.BremsMonoEnergySpectrum}), 
the photon production rate is analytically given by
\begin{eqnarray}
N(k) & = A\int^{\epsilon_{\mathrm{max}}}_{\epsilon_{\mathrm{min}}} N(\epsilon_\mathrm{ini},k)
f(\epsilon_\mathrm{ini}) d \epsilon_\mathrm{ini} \nonumber\\
&= A\int^{\epsilon_{\mathrm{max}}}_{\epsilon_{\mathrm{min}}} d \epsilon_\mathrm{ini}
f(\epsilon_\mathrm{ini}) \int^{\epsilon_{\mathrm{ini}}}_{k+1} d \epsilon
\frac{d\sigma}{d k} v(\epsilon) \left|\frac{d\epsilon}{dt}\right|^{-1}.
\label{eq.PowBremsThick}
\end{eqnarray}
This prediction is again show on the figure. 
The simulation agrees within $\sim 20$\% with the analytic calculation 
via eq.(\ref{eq.PowBremsThick}). 

As can be seen from figure \ref{fig.PowThickSim}, 
the photon spectra emitted by power-law distributed electrons 
under the thick-target condition again exhibit power-law shape up to $\sim 10$ MeV,  
like in the case of the thin-target emission (figure \ref{fig.powerThinSpectrum}). 
The obtained photon index is $\gamma \simeq 1.4$ for $\delta\simeq0.8$ 
and $\gamma \simeq 1.5$ for $\delta\simeq 1.5$.  
In contrast, the hardest portion of the spectra exhibits rather opposite trend, 
and cannot become as hard as implied by equation.(\ref{eq.Brown}),  
since the maximum energy of electrons decreases as they propagate deeper 
through the thick target. 

Figure \ref{fig.thinthick} (right) is the same comparison
as figure \ref{fig.thinthick} (left),
but for the thick-target condition.
Photon indicies in this case are generally smaller compared with thin target emission, 
because of the loss-flattening effect (section \ref{sec.introduction}).
In addition, $\gamma$ depends more weakly on $\delta$ 
than in the thin-target case.

\clearpage

%%%%%%%%%%%%%%%%%%%%%%%%%%%
\section{Solar Simulation}\label{sec.SolarSimulation}
%%%%%%%%%%%%%%%%%%%%%%%%%%%

%--------------------------------------------------
\subsection{A more realistic model atmosphere}
%--------------------------------------------------

In the previous section, we treated the solar atmosphere 
as a uniform gas slab. 
This is basically reasonable, because the energy losses of electrons 
and photons are primary determined by the column density of matter 
along its path. 
The actual solar atmosphere, with the strong vertical density gradient, 
could be transformed into such a uniform zone, 
as long as all the particles are traveling vertically.  
However, particle trajectories with significant transverse components 
can no longer be conserved by 
such a transform, and hence we need to construct a more realistic 
solar model taking into account its vertical density gradients. 
We here refer to ``Harvard-Smithsonian reference atmosphere'' \citep{Gingerich1971}
for the solar mass-density profile,  
and approximate this numerical table analytically by an 
exponential function as 
\begin{equation}
\rho(z) = 3.19\times10^{-7}\exp\left(-\frac{z}{h}\right).  
\label{eq.harvardDensity}
\end{equation}
Here, $z$ is the vertical coordinate measured from the photosphere, 
$\rho(z)$ is the mass density in the unit of g cm$^{-3}$, 
and $h$ is the scale height which is $\sim 400$ km for $z<0$ 
(the solar interior) and $\sim 110$ km for $z>0$ (the coronal region). 

We construct our new solar atmosphere model as a series of boxes (or ``zones'') 
stacked in the vertical direction. 
Individual zones are defined to have different densities 
but to share approximately the same column density, 
and the density within each of them is assumed to be uniform. 
Each zone is assumed to have a lateral extent of $2\times 10^4$ km, 
to simulate the cross section of a typical magnetic loop.  
On this transverse scale, the curvature of the solar surface is negligible. 
As illustrated in figure \ref{fig.zones}, 
we divide the coronal ($z>0$) region into 5 zones, 
each having approximately the same column density of 0.70 g cm$^{-2}$. 

According to the Harvard Smithonian model, 
the total column density above $z = 0$ is 3.5 g cm$^{-2}$,  
which corresponds to a depth of $\sim 100$ km in our simple atmosphere model. 
Also, the sum of the 5 coronal zones in our new model gives an 
overall column density of 3.5 g cm$^{-2}$,  
 thus faithfully representing the actual solar atmosphere for $z>0$. 
While traversing each zone, a 100 MeV electron loses $\sim 15$ MeV in total , 
of which only $\sim 20$ \% is in radiation whle the rest is in Coulombic loss. 

In simulating the solar interior ($z<0$),  we use 10 thicker zones 
having 3.5 g cm$^{-2}$ each;  
again, this corresponds to a depth of $\sim 100$ km in our simple atmosphere model. 
With the overall column density of 35 g cm$^{-2}$ below $z=0$, 
our new model goes down to a depth of $\sim 500$ km, 
below the solar photosphere  in comparison with equation (\ref{eq.harvardDensity}). 
This depth is considered sufficient, because a 100 MeV electron, 
injected at the top of our model, 
would lose $\sim 100$ \% of its initial energy by the time it reaches the bottom of the deepest zone. Furthermore, the thicker 10 zones, when summed up, 
has a Compton optical depth of $\sim 10$, for a vertically precipitating photons; 
few photons would penetrate deeper. 

Although our ``solar atmosphere'' is comprised of pure hydrogen,  
 the actual solar atmosphere contains helium to $\sim 10\%$ by number,  
 or $\sim 25\%$ by mass. 
Since the ionization loss of electrons is proportional to the electron number density, the helium would decrease the ionization loss per unit mass column density by $\sim 10\%$.  
Bremsstrahlung per unit mass should remain the same, 
because a helium atom has 4 times higher mass than a proton 
but 4 times higher bremsstrahlung cross section at the same time. 
The Compton scattering depends only on the number of electrons,  
and hence would decrease by $\sim 10\%$ if helium was included.

%---------------------------------
\subsection{vertical injection}
%---------------------------------

\begin{figure}[htbp]
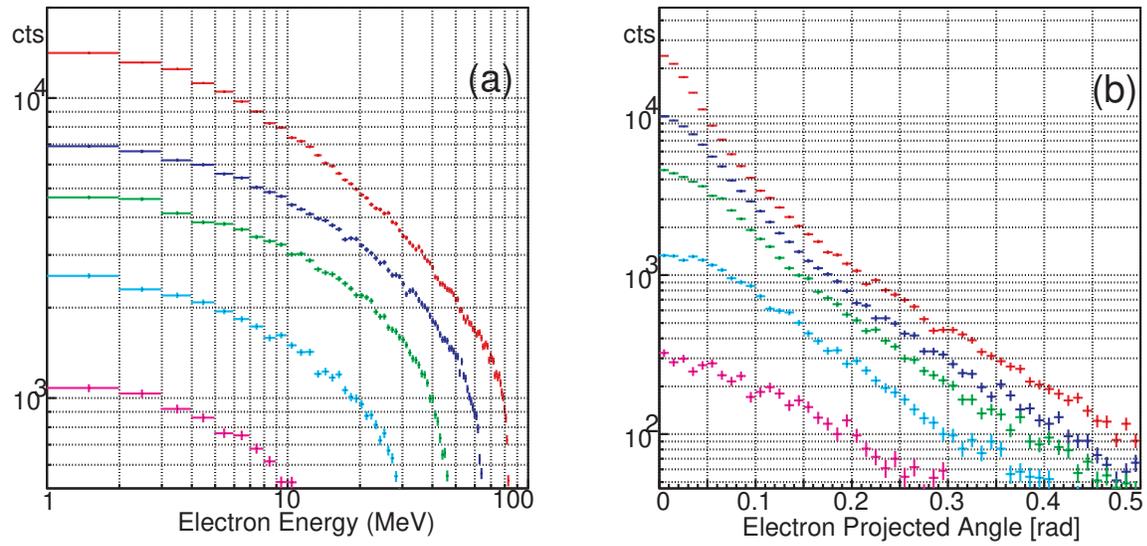

\begin{center}
\FigureFile(8cm,8cm){v9.921to100MeVG1.2e1000000ElectronSpectrum.eps}
\FigureFile(8cm,8cm){v9.921to100MeVG1.2e1000000ElectronProjAngDist.eps}
\caption{The same as figure \ref{fig.monoElectronSpectrum}, 
but when the initial electrons are distributed between 1 and 100 MeV 
with a power-law spectrum of photon index $\delta = 1.2$. 
Red, blue, green, cyan, and magenta indicate depths of 100, 200, 300, 
 400 and 500 km, respectively.}
\label{fig.powElectronSpectrum}
\end{center}
\end{figure}

\begin{figure}[htbp]
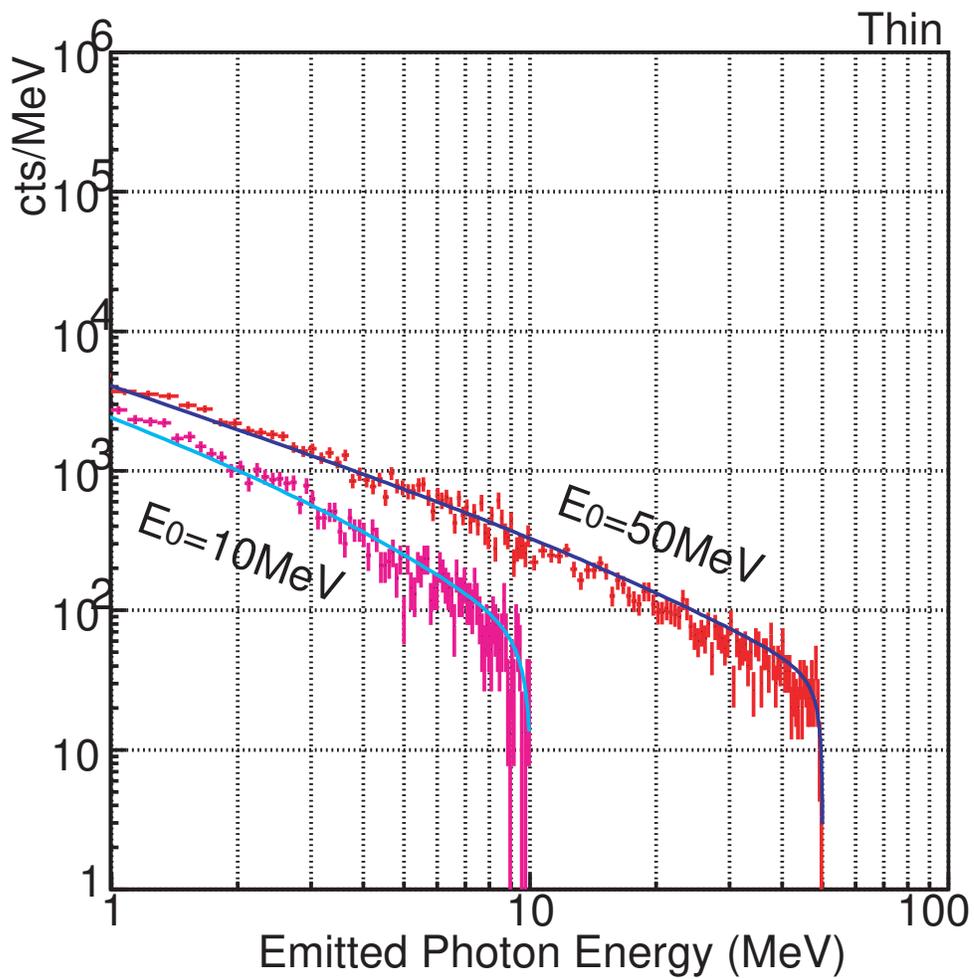

\begin{center}
\FigureFile(14cm,14cm){thinTargetEmission2.eps}
\caption{Thin-target bremsstrahlung photon spectra, produced in a hydrogen gas 
slab (the thin atmosphere model) by mono-energetic electrons with 
kinetic energies of 10 MeV (magenta) and 50 MeV (red).  
Crosses show the {\sc Geant4} simulation results, 
using the thin atmosphere model geometry, 
while solid curves refer to the analytic expression by \citet{KochandMotz1959}}
\label{fig.BremsMonoEnergySpectrum}
\end{center}
\end{figure}

\begin{figure}[htbp]
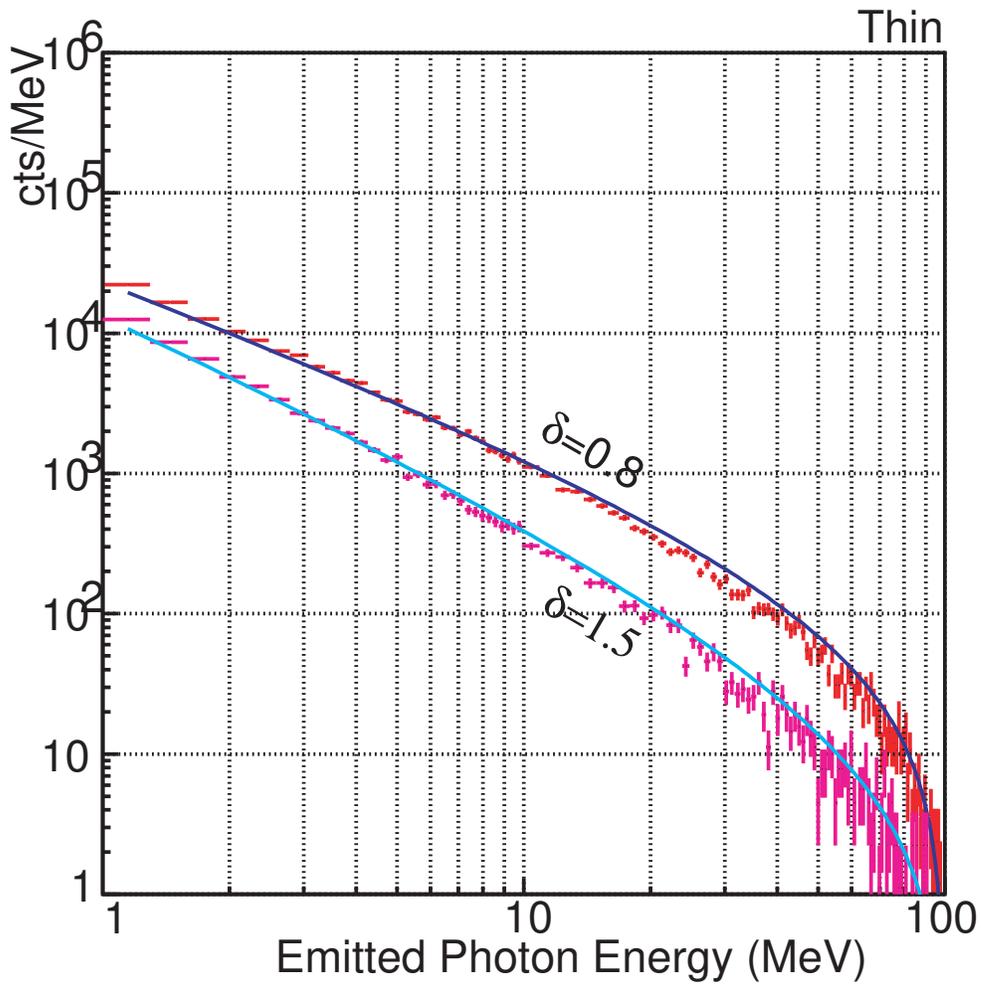

\begin{center}
\FigureFile(14cm,14cm){thinTargetPowrLawEmission2.eps}
\caption{Thin-target bremsstrahlung spectra produced by incident electrons
having a power-law index of $\delta = 0.8$ and 1.5, distributed over the 1-100 MeV range.
Crosses show the {\sc Geant4} simulation results 
using the thin atmosphere model geometry, 
while solid curves refer to the analytic expression by equation (\ref{eq.PowBremsThin})
}
\label{fig.powerThinSpectrum}
\end{center}
\end{figure}

\begin{figure}[htbp]
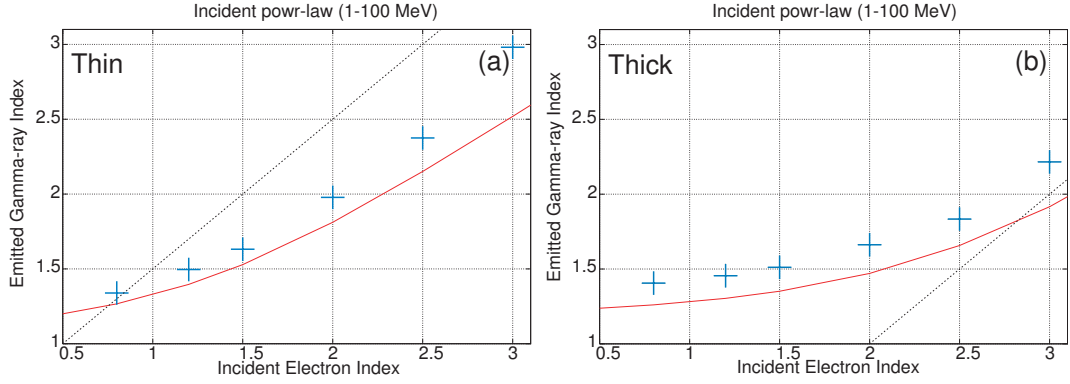

\begin{center}
\FigureFile(7cm,7cm){ThinpowRebinCollected5.eps}
\FigureFile(7cm,7cm){ThickpowCollected5.eps}
\caption{(left) Relation between the index $\delta$ of power-law distributed (1-100 MeV) electrons and the resulting thin-target bremsstrahlung photon index $\gamma$.
 The red lines is the calculation from equation (\ref{eq.PowBremsThin}), 
blue crosses show the results of {\sc Geant4} simulations, 
and the black line refers to the non-relativistic formula represented by equation (\ref{eq.BrownThin}). 
The photon index refer to the 1-10 MeV range.  
(right) 
The same as left figure, but for the thick target emission. 
The analytical results (red) is calculated from equation (\ref{eq.PowBremsThick}), 
while the blue line represents the non-relativistic relation of in equation (\ref{eq.Brown}). 
}
\label{fig.thinthick}
\end{center}
\end{figure}

\begin{figure}[htbp]
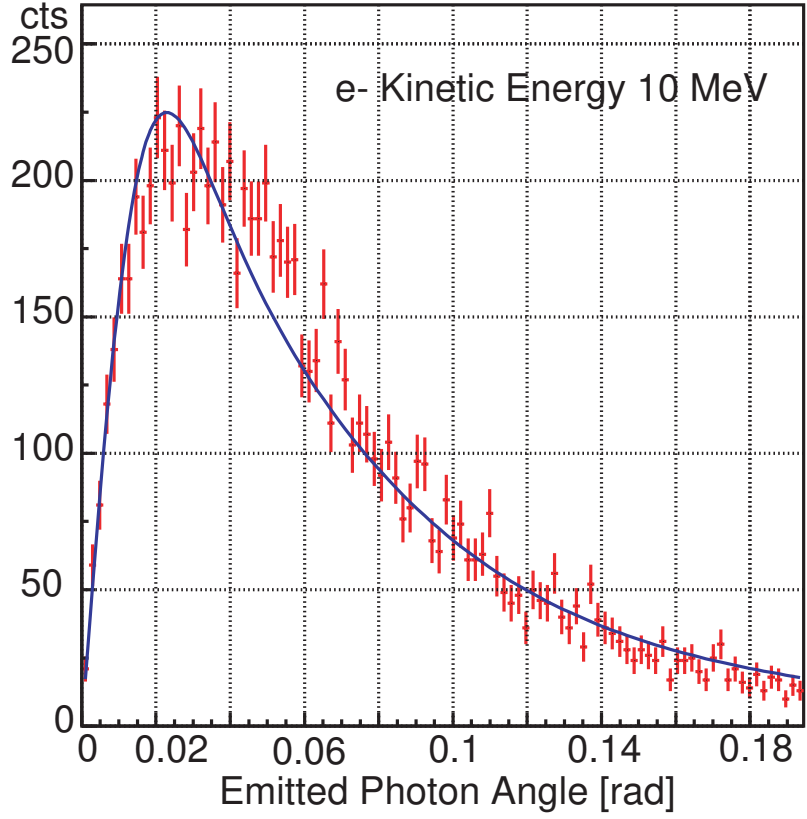

\begin{center}
\FigureFile(12cm,12cm){10MeVe100000000argfitkm2bs.so.eps}
\caption{
Angular distributions (per unit polar angle measured form the electron motion) 
of thin-target bremsstrahlung photons,
produced by electrons with the kinetic energies of 10 MeV.
Crosses represent the Monte Carlo simulations,
while solid curves are analytic expressions by equation (\ref{KM2BS}).}
\label{fig.BremsAngDistValid}
\end{center}
\end{figure}

To examine the flare photon spectra, 
we vertically injected $10^6$ electrons into our solar atmospheric model from a height of 
$z = 10,000$ km, and tracked them from that height to $-500$ km, 
as they lose energy via Coulomb scattering and emit bremsstrahlung photons. 
The electrons are assumed to be distributed initially in the 1-100 MeV range 
with a power-law index $\delta=1.2$. 
The electrons, initially moving in the vertical direction, 
gets gradually deflected as indicated in figure \ref{fig.powElectronSpectrum}; 
in actual configuration, this corresponds to pitch-angle scattering. 
Nevertheless,  as represented by figure \ref{fig.powElectronSpectrum} in red, 
still 98 \% of the electrons are contained within $\sim 5^{\circ}$ 
of the vertical direction when they have reached the photosphere. 
Therefore, the vertical injection employed here is valid 
as long as the magnetic fields are sufficiently normal to 
the solar photosphere,   
the electrons' pitch angle is initially concentrated near $\sim 0^{\circ}$,  
and non-Coulombic (e.g., magnetohydro-dynamic,
or due to magnetic mirroring) pitch-angle scatterings 
can be neglected. 

The bremsstrahlung photons emitted by the almost 
vertically moving electrons are strongly forward collimated, 
as indicated by figure \ref{fig.BremsAngDistValid}. 
Therefore, most of them are expected to enter the atmosphere
and undergo Compton scattering or photoabsorption.
Although some of these photons will die,
others will be scattered back to escape out of the photosphere. 
We collected these outcoming photons at different viewing angles $\beta$ (figure \ref{fig.coordinate}),  
and averaged the results over $0\leq\varphi<2\pi$. 
Figure \ref{fig.ComptonGamma0Deg} shows the photon spectra obtained in this way for equal intervals in $\cos\beta$ (hence over an equal solid angle).   

In figure \ref{fig.ComptonGamma0Deg}, spectra with small $\beta$, simulating disk-center flares, 
exhibit a clear ``knee'', which represents the effect of single Compton scatterings. 
In the case of $\beta\sim 0$ (blue in figure \ref{fig.ComptonGamma0Deg}), 
the knee energy becomes $\sim 1/2\ m_e c^2$, 
because of the strong energy degradation in the single Compton back-scatterings. 
Photons with energies above the knee result 
either directly from the bremsstrahlung source, 
or through multiple Compton scatterings   
with a relatively small energy loss in each step. 
The spectrum in this region has a much steeper slope than in the region below the knee. 

As $\beta$ gets larger in figure \ref{fig.ComptonGamma0Deg}, 
the knee energy increases, 
because the angle needed in the single Compton scattering decreases. 
Nevertheless, the spectrum remains much softer 
(e.g., $\gamma\sim 3$ in the 0.2-0.6 MeV range) than is expected 
when the Compton effects are not taken into account ($\gamma\sim 1.3$; figure \ref{fig.thinthick}b). 
The observed photon flux below the knee energy decreases 
as $\beta$ increases and approach $90^{\circ}$, 
due to the same limb darkening effect as in visible light.

%----------------------------
\subsection{Slant injection}
%----------------------------

As we have confirmed in figure \ref{fig.ComptonGamma0Deg}, 
bremsstrahlung photons back-scattered  
from the solar photosphere exhibit very steep spectra 
as a result of the Compton degradation.  
In order to explain the origin of the observed flare spectra often extending beyond 1 MeV, 
we must then considered photons which reach us after Compton scatterings  
with much smaller angles.  
For this purpose, we have to consider electrons penetrating into the atmosphere with shallow angles (i.e., large $\alpha$), namely  ``slant injection''. 

We then examined the case of slant injection, by shooting  
$10^6$ power-law electrons (in the 1-100 MeV energy range),
into our realistic solar model.
Specifically, we repeated the {\sc Geant4} simulation under the same condition, but assuming this time an injection angle of $\alpha = 80^{\circ}$ (figure \ref{fig.coordinate}),  
instead of the vertical injection ($\alpha = 0^{\circ}$) considered 
in the previous subsection. 
We now inject the electrons from a height of 100 km 
(i.e., neglecting the uppermost two zones in figure \ref{fig.zones}),
to avoid them escaping sideways before arriving at thicker regions.
Figure \ref{fig.ComptonGamma80Deg} shows the results of this simulation 
for the same three intervals of $\beta$ as figure \ref{fig.ComptonGamma0Deg},
again averaged over the observer's azimuth angle $\varphi$. 
The spectrum becomes considerably harder than in the case of 
vertical electron injection, 
and extends much beyond $\sim 1$ MeV 
when $\beta$ is rather large ($0.1 < \cos\beta < 0.2$).  
In this case, electrons make grazing incidence on the atmosphere, 
and some of their bremsstrahlung photons get nearly forward scattered to leave the zone 
with a small energy loss, thus producing the hard power-law continuum.

In this way, we have Monte-Carlo simulated slant
injections for a family of electron spectra with various indicies
$\delta$ (again distributed over 1--100 MeV),
assuming two representative injection angles
of $\alpha=60^\circ$ and $\alpha=80^\circ$. 
Figure \ref{fig.DeltaGamma} summarizes the photon index $\gamma$ of
1--10 MeV gamma-rays, collected over the most
favorable viewing angle of $0.1 < \cos \beta < 0.2$
(i.e., rather grazing to the solar photosphere)
and integrated over the observer's azimuth.
Thus, even with the extreme assumption of $\alpha=80 ^\circ$,
the simulated gamma-ray spectra become significantly
softer (typically by $\sim 1$ in $\gamma$)
than the original bremsstrahlung spectra
before the photons are Compton scattered
(reproduced in figure \ref{fig.DeltaGamma} in blue).
Moreover, $\gamma$ further increases by $\sim 0.5$
as $\alpha$ decreases from $80^\circ$ to $60^\circ$.
These results imply
that the Compton scattering significantly soften
the emergent photon spectra,
of which the effect depends sensitively on $\alpha$ and $\beta$.

An interesting prediction of figure \ref{fig.DeltaGamma} is that 
the observed gamma-ray photon index should appear 
in a relatively narrow range of $\gamma=1.7-2.6$, 
regardless of the electron index $\delta$. 
Of course, we would observe considerably larger values of $\gamma$ 
when either $\alpha$ or $\beta$ is rather small. 
However, such a spectral component would not be easily detected 
in the MeV energy region. 

Although our simulations of the slant injection neglected 
the uppermost two coronal zones, their effects may not be necessarily 
negligible, 
because the large injection angle $\alpha = 80^\circ$ (or $60^\circ$) 
effectively increases the electron path length in each zone 
by $(\cos 80^\circ)^{-1} = 5.8$ [or $(\cos 60^\circ)^{-1} = 2.0$] times.  
Since the electron energy losses in these two regions, 
if properly taken into account, would reduce the maximum electron energies 
while flatten the low-energy electron slope (figure 3), 
the emergent photon flux above $\sim$ 1 MeV would further be suppressed 
if the electron injection in actual flares takes place at a height much 
exceeding the assumed 100 km. 
In other word, our simplification employed in simulating the slant injection 
makes the case more conservative.

\clearpage
%%%%%%%%%%%%%%%%%%%%%
\section{Discussion}
%%%%%%%%%%%%%%%%%%%%%

%---------------------------------------
\subsection{Summary of simulations}
%---------------------------------------
In order to investigate the effects of Compton scatterings 
on the solar gamma-ray spectra, 
we have conducted Monte Carlo simulations of energetic electrons, 
impinging on the solar atmosphere,   
using {\sc Geant4} as our basic toolkit.  
We modeled the solar atmosphere as vertically stacked parallel hydrogen zones, 
each assumed to be uniform.
The density gradient was expressed by assigning higher densities to lower zones. 
We injected electrons which are power-law distributed over the 1-100 MeV range,   changing their spectral index $\delta$ and incident angle $\alpha$. 
The results of the present simulations can be summarized as follows. 
\begin{enumerate}
\setlength{\itemsep}{-1mm}
\item{The observed photons hardly reach MeV energies 
if the electrons are injected vertically,  
because the bremsstrahlung-produced gamma-rays must be Compton back-scattered  
to reach the observer, and hence their energies are subject to strong Compton 
degradation.}
\item{The emergent gamma-ray spectrum, integrated over the observer's azimuth, 
gradually hardens as the injection angle $\alpha$ increases 
(becoming more ``slant''), 
and also as the observer's polar angle $\beta$ gets larger. 
}
\item{As long as $\alpha$ and $\beta$ are both relatively large, 
the azimuthally-averaged 1-10 MeV gamma-ray spectra exhibit 
$\gamma=1.7\sim2.6$ over a relatively wide range of $\delta$.}
\end{enumerate}
These results generally agree with, and further extend,
the previous works by, e.g., \citet{BaiRamaty1978} and \citet{Kontar2006}.

%-----------------------------------------------------
\subsection{Effects ignored in the simulation}
%-----------------------------------------------------
There are some effects which are not considered in our simulation.  
Among them, those of the secondary electrons and helium 
were already described in section 2.2 and section 3.1, respectively. 
As the electron energy increases, their synchrotron energy loss 
becomes significant. 
However, in the chromospheric region in which we are interested, 
the magnetic field intensity is typically $\sim 100$ G, 
and hence 
the rate of synchrotron loss of 100 MeV electrons is only $\sim 1$ \% 
of their Coulomb loss, and only $\sim 5$ \% of their bremsstrahlung loss.  
This is because the Coulomb and bremsstrahlung losses 
are rather large due to the high matter density. 
Thus, the synchrotron process can be neglected. 

We also ignored the inverse Compton process between flare-accelerated electrons 
and ambient photons, 
because electrons with the maximum energy assumed, 100 MeV, 
can produce at most hard X-rays up to $\sim 100$ keV 
by scattering off the solar visible photons. 
However, if the electrons have a very flat spectrum  
well extending to energies above 350 MeV, 
 the inverse Compton process may become a key process 
in creating gamma-ray photons. 

The size of the world used in our simulation, 
10000 km $\times$ 10000 km $\times$ 20000 km,   
simulates a typical size of a foot point of a magnetic tube. 
The number of photons which escaped from the boundaries 
below the photosphere is less than $10^{-4}$ of that of produced photons, 
and energies of these escaped photons are less than 100 keV each. 
Hence we ignored those photons in the simulation. 
Possible effects of the relatively low heights of our slant injection was 
already considered in subsection 3.3.

In the present simulation, we truncated the power-law electron spectrum 
at 100 MeV. In order to examine whether this abrupt cutoff produces any 
artifact, we repeated some of the simulations by changing the maximum 
cutoff energy to 1 GeV. q
Then we have found that $\gamma$ decreases only slightly, 
typically 0.1 or less,  
if $\delta$ is larger than 2. 

Finally, a separate gamma-ray component could be produced
without the Compoton effect, simply if the electron spectrum
consists of two power-law components with different slopes.
Even so, the good agreement between the predicted range of
power-law indices and the {\em Yohkoh} data suggests that the
Compton scattering is playing a significant role (see the next section).

%----------------------------------------------------------------------------
\subsection{Comparison with observations}
%----------------------------------------------------------------------------
In order to evaluate the significance of the results of our simulation, 
let us compare them with actual observations of hard X-ray 
and gamma-ray emission from solar flares. 
For that purpose, 
we have also to relate the simulation parameters $\alpha$ and $\beta$ 
to observational parameters,   
including pitch-angle distributions of electrons, 
the tilt of magnetic field lines, 
and the flare longitude on the solar disk. 
As the observational data, 
we utilize 40 flares that show significant gamma-ray emissions, 
selected out of 2788 X-ray flares observed from 1991 October to 2001 December 
by {Yohkoh} \citep{Yukari2005PASJ}. 
{Yohkoh}, the Japanese solar satellite launched in 1991, 
was able to take four-color hard X-ray flare images 
with the Hard X-ray Telescope(HXT) \citep{Kosugi1991,Kosugi1992} 
and perform spectro-photometry over the $0.2-30$ MeV range 
with the Gamma-Ray Spectrometer (GRS) \citep{Yoshimori1991}. 

%-----------------------------------------------------------
\subsubsection{Vertical magnetic fields}
%----------------------------------------------------------
As the simplest case, we may consider a condition
where the magnetic field lines are perpendicular to the photosphere. 
In this case, we can identify $\alpha$ 
with the pitch-angle of an electron, 
and the effect of electron gyration around the magnetic field lines is 
fully represented by our photon collection method which takes an average 
over $\varphi$. 
If the electrons have a narrow pitch-angle distribution around 0, 
the strong Compton degradation, 
described in section 3.2 
and shown in figure \ref{fig.ComptonGamma0Deg}, 
will make the observed photon spectrum very steep, 
and will prevent it from reaching MeV energies. 
If, on the other hand, the pitch-angle distribution is rather broad, 
electrons with larger values of $\alpha$ will emit harder spectra, 
as represented by figure \ref{fig.ComptonGamma80Deg}. 
Conversely, as long as the field lines are close to vertical, 
flares with significant gamma-ray emission must involve broad pitch-angle 
distributions of electrons. 

Analyzing the 40 gamma-ray flares detected by {Yohkoh}, 
\citet{Yukari2005PASJ} found that their gamma-ray (typically 1 MeV) to hard X-ray 
($\sim$ 70 keV) flux ratios scatter largely (by an order of magnitude), 
with no correlations to their hard X-ray spectral slopes. 
Hence they conclude that there is a hidden parameter which causes the gamma-ray flux to vary significantly relative to the hard X-ray extrapolation, 
in such a way that the gamma-ray component often appears as a separate spectral 
hard tail (see their figure \ref{fig.thinthick}). 
Given the present results, this parameter is very likely to be 
the electron pitch-angle distribution, 
with broader distributions yielding higher gamma-ray 
to hard X-ray flux ratios. 

In addition to the electron injection angle $\alpha$, 
the viewing angle $\beta$ has been confirmed to strongly affect 
the flare gamma-ray spectra (section \ref{sec.SolarSimulation}). 
Assuming again the vertical field geometry, 
we can identify $\beta$ with the angle $\theta_v$ which our line-of-sight 
to the flare makes relative to the normal to the photosphere. 
Hereafter, we simply call $\theta_v$ ``flare-viewing angle''. 
Using the flare longitude $l$ and latitude $b$, 
it can be calculated as 
\begin{equation}
\theta_v = \cos^{-1} ( \cos b \,\cos l ). 
\label{eq.theta_v}
\end{equation} 

To examine the effect of $\beta$, 
we plotted in figure \ref{fig.longitude} the gamma-ray to hard X-ray 
hardness ratio  of the 40 {Yohkoh} flares, 
as a function of $\theta_v$ calculated via equation (\ref{eq.theta_v}). 
It is similar to figure 7 of \citet{Yukari2005PASJ}, 
but differs from it in that $\theta_v$ is used instead of $|l|$. 
As predicted by our simulation, 
the hardness ratio clearly increases with $\theta_v$; 
this effect was already known as ``gamma-ray limb brightening'' 
in solar flares using {\it SMM} data \citep{Vestrand1987}. 
\citet{Vestrand1987} interpreted this effect 
in terms of the radiation anisotropy:
supposing that the electrons have a pitch-angle distribution
which increases with angle from the outward normal,
then higher-energy continua are expected to be
more strongly beamed parallel to the photosphere,
because harder bremsstrahlung photons are
more strongly forward-beamed than softer ones
even when emitted by electrons of the same energy.
We presume that this mechanism can explain
at least part of the observed limb-hardening effects,
while the Compton back-scattering process
must be enhancing them.
In particular, our scenario can explain limb hardening
even when the electrons are mostly directed downwards.

For a more quantitative comparison of our simulation 
with the {Yohkoh} observations, 
let us limit ourselves to limb flares which have absolute longitudes 
larger than $60^{\circ}$, 
because gamma-ray data of disk flares are of poor photon statistics 
than those of limb flares 
as is clear from figure \ref{fig.longitude},  
and because these disk flares are more contaminated by nuclear gamma-ray lines 
\citep{Yukari2002} which are unrelated to the accelerated electrons. 
Figure \ref{fig.color_color2} shows a scatter plot between two hardness ratios 
 of the 40 sample flares, one calculated in the hard X-ray range (57---93 keV vs 33---57 keV) 
while the other in the gamma-ray range (1.43---6.21 MeV vs 0.22---1.43 MeV). 
In this ``color-color'' plot, 
the dashed line represents the condition that the two colors imply the same 
photon index $\gamma$, of which values are given in the figure. 
The hard X-ray (or gamma-ray) photon index of a flare can be found 
by drawing a vertical (or horizontal) line from the data point, 
and read its intersection with the dashed line.  

From figure \ref{fig.color_color2}, 
we can see that the gamma-ray spectral index of the 40 flares is 
distributed in a typical range of 1.7-2.5. 
This agrees very well with the prediction by our Monte-Carlo simulations 
(figure 13; subsection 3.3), 
on condition that the electrons in these flares have broad pitch-angle 
distributions beyond $\alpha\sim 60^\circ$. 
Further, in figure \ref{fig.color_color2}, 
a majority of flares are distributed 
to the left side of the single power-law locus.  
Such a concave-shaped wide-band spectrum can be naturally explained 
as a consequence of a broad pitch-angle distribution of electrons,  
because electrons with small pitch angles are expected 
to emit steep-spectrum photons, 
while those with large pitch angles emit much harder spectra. 
A superposition of these spectral components with different slopes 
will result naturally in a concave spectrum just as is observed. 
Thus, as long as the magnetic lines are assumed perpendicular 
to the photosphere 
and the flare viewing angle $\theta_v$ is large, 
flares with significant gamma-ray signals are inferred 
to have rather broad pitch-angle distributions. 

%-----------------------------------------------------------
\subsubsection{Slant magnetic fields}
%----------------------------------------------------------

As a more general case, we may briefly consider tilted magnetic-field conditions. 
In this case, it is somewhat difficult to directly compare the results 
of our simulation with the observational data, 
because the electron gyration can no longer be faithfully emulated by our method of taking an average over $\varphi$. 
Nevertheless, we may employ the present results to predict some aspects of that case in a qualitative way. 

Let us first consider limb flares as an extreme case. 
In this case, a field-line tilt on the sky plane is considered 
to have little effects, because the most efficient gamma-ray emission 
(as seen from the observer) occurs over the gyration phase 
wherein the electrons are approaching us, and this phase is the same 
even the filed lines are tilted in that way. 
In contrast, a field-line tilt along our line-of-site is expected 
to have a prominent effect: 
if the field lines are tilted away from us, 
the observed spectra tend to reach higher energies, e.g. a few MeV, 
even if the electrons have a rather narrow pitch-angle distribution. 
If the lines are tilted toward us, 
the gamma-ray spectra are expected to soften, on the contrary. 
Then, the observed scatter in the gamma-ray to hard X-ray intensity ratio 
for limb flares (figure \ref{fig.longitude}, with $\theta_v \gtrsim 60^{\circ}$) may be contributed 
by the field-line tilts, in addition to the pitch-angle distribution. 

The other extreme case is a disk-centered flare, 
where the field-line tilt is specified only by its angle $\theta_t$ 
with respect to the photosphere normal (i.e., our line-of-sight) 
without depending on the tilt direction. 
Then, an electron with a fixed pitch angle, $\alpha_0$ is expected 
to sweep a range of injection angle, 
from $\alpha_-=|\alpha_0 - \theta_t|$ to $\alpha_+=\alpha_0 + \theta_t$. 
Therefore, a larger value of $\theta_t$ is favorable 
in delivering strong gamma-ray signals to the observer, 
even though $\beta$ is still limited to the unfavorable value of $\sim 90^\circ$. 
If $\alpha_0$ and $\theta_t$ are both rather large so that $\alpha_+ > 90^\circ$, the electron is implied to be approaching us over a certain phase of its gyration, thus allowing a considerable fraction of the bremsstrahlung photons to directly reach us without being Compton scattered. 
Therefore, a disk flare may become a strong gamma-ray source 
under a limited condition, that the electrons have a broad pitch angle 
and the magnetic field lines are significantly tilted. 

%--------------------------------------------------------
\subsubsection{Case of a  broad pitch-angle distribution}
%--------------------------------------------------------

Now that  broad pitch-angle distributions of accelerated electrons 
are considered to be an important element of 
observing intense gamma-ray emission,
it may be imperative to simulate such a case,
and compare the  results with gamma-ray spectra from actual solar flares.
Therefore, we repeated our solar simulation, 
assuming that  the electrons have a continuous incident-angle distribution as  
\begin{equation}
\propto \frac{3}{2}\,(1-\cos^2\alpha) ~~,
\label{eq.DermerRamaty1986}
\end{equation}
which corresponds to  $M_2(\cos\alpha)$ in \citet{Dermer1986} 
except for covering only the $0 \leq \cos\alpha \leq 1$ range.
When the magnetic fields are vertical,
equation (\ref{eq.DermerRamaty1986}) becomes 
identical to the pitch-angle distribution of the accelerated electrons.
Under this angular distribution,
we injected $10^7$ electrons 
in the same way as in section~\ref{sec.SolarSimulation},
assuming the  incident electron spectrum to
have a canonical  index of $\delta=2.0$,
and  to extend again over 1--100 MeV.
Collecting the produced gamma-rays over a viewing angle range
of $0.1 \leq \cos\beta \leq 0.2$,
and convolving the results with an approximate response
of the Yohkoh GRS \citep{Yoshimori1991},
we have obtained the spectrum shown in figure~\ref{fig.grs_sim_comp} in red.

To compare with this simulation, 
we have chosen a GOES X3.7 class flare of 1998  November 22, 06h37m UT,
which is one of the gamma-ray brightest events among 
the 40  Yohkoh flares  \citep{Yukari2005PASJ}.
Since this is an extreme limb flare located at  S32W90,
our viewing angle $\beta$ should be close to $90^\circ$,
in agreement with the simulation condition.
Actually, the HXT highest-band images of this flare reveal 
intense hard X-ray emission from a pair of magnetic loop foot points,
both located right on the solar west limb  \citep{Yukari2002}. 
The GRS data, 
accumulated over a period of 06:37:22--06:40:22,
are shown in figure \ref{fig.grs_sim_comp} in blue,
after subtracting background (accumulated for 06:56:14--06:59:02),
but without removing the detector response.
Thus, the spectrum is featureless, 
and the 2.23 MeV neutron capture line is undetectable
with its equivalent width being $<10$ keV  \citep{Yukari2002}.
Therefore, this flare is considered to be a
typical ``electron dominated''  event.

In figure~\ref{fig.grs_sim_comp}, 
the simulation and the actual data agree reasonably well,
at least over the 1--10 MeV band.
When the GRS response is considered,
these spectra  can be approximated both by 
$\Gamma = 2.0 \pm 0.1$ \citep{Yukari2002},
which is typical in figure~\ref{fig.color_color2}.
Incidentally, we would obtain from figure~\ref{fig.DeltaGamma}
$\Gamma \sim 1.7$ for the assumed $\delta=2.0$,
if we were observing direct thick-target emission.
We should instead find $\Gamma=2.2$ (for $\alpha=80^\circ$)
to $\Gamma=2.5$ (for $\alpha=60^\circ$),
if  the Compton back-scattering dominates.
Therefore, the values of $\Gamma \sim 2.0$
found in figure~\ref{fig.grs_sim_comp}
are in between these two cases.
Indeed, an inspection  of the Monte-Carlo photons revealed
that $\sim 60\%$ of them are Compton-scattered gamma-rays,
while the remaining $\sim 40\%$ are direct bremsstrahlung events.
Given the good agreement between the two data sets,
it is possible that the actual gamma-rays from the November 22 flare 
are also a mixture of the direct and scattered photons.

%===========================
\section{Conclusion}
%===========================

In order to quantitatively estimate the spectra of energetic electrons 
generated in solar flares based on the observed gamma-ray spectra, 
we numerically studied the electron transport and gamma-ray emission 
in the solar atmosphere. 
As elementary processes, we mainly considered Coulomb scattering, 
bremsstrahlung and Compton scattering. 
We modeled the solar atmosphere with a vertical stack of parallel zones 
with different densities, and neglected the magnetic fields. 

As a result of our simulation, 
we have found that the thick-target gamma-ray spectra emitted 
from magnetic loop foot-point can hardly reach MeV energies, 
because of heavy Compton degradation, 
unless the electrons make rather grazing angles to the photosphere 
and we are observing forward-scattered gamma-rays. 
When this Compton effect is taken into account, the traditional $\delta$ vs. $\delta$ relation is drastically modified. 
The pitch-angle distribution of flare-accelerated electrons 
are suggested to be broad and extend to large angles for those flares 
of which the emitted photon spectra reach MeV energies. 
Actually, a  gamma-ray spectrum simulated
under electron injection with a  broad pitch angle distribution
reproduced reasonably well the Yohkoh spectrum of the 1998 November 22 flare.

In terms of actual flares,
the above results can be translated into the following conclusions.
\begin{enumerate}
\item{When the magnetic field lines are perpendicular to the photosphere,
and the electrons have a narrow pitch-angle distribution around 0,
the gamma-ray spectra are very soft and cannot reach MeV energies.}
\item{Even in the same field configuration,
the gamma-ray spectra become harder
if the electrons have larger pitch angles.
Therefore, the electrons producing gamma-rays are inferred to have relatively
large pitch angles.}
\item{Limb flares are expected to have harder spectra than disk flares,
because of their smaller angles of Compton scattering.
This agrees with the observed fact that the gamma-ray to hard X-ray
intensity ratio increases as the flare longitude gets closer
to the solar limb.}
\item{In case of a broad pitch angle distribution,
electrons with small pitch angles emit steep hard X-rays,
while those with large angles emit photons with flatter spectra.
This can explain another {Yohkoh} result, that the gamma-ray slope
tends to be flatter than the contemporaneous hard X-ray slope.}
\item{Under the conditions of $\delta = 1-3$ and $\alpha > 60^{\circ}$, 
the values of $\gamma$ obtained in our simulations in the 1--10 MeV range
are concentrated over 1.7--2.5.
This agrees again with the Yohkoh observations.}
\item{
A disk flare may become a strong gamma-ray source under a limited condition, that the electrons have a broad pitch-angle and the magnetic field lines are significantly tilted. 
}
\item{
The  gamma-ray spectra of actual solar flares are expected to be
an appropriate mixture of  the direct and Compton-scattered photons.}
\end{enumerate}

%\bibliographystyle{apj}
%\bibliography{Dron}
% Reference 

%%%%%%%%%%%%%%%
% Figures
%%%%%%%%%%%%%%%

\clearpage

\clearpage

\clearpage

\clearpage

\clearpage

\clearpage

\clearpage

\clearpage
\begin{figure}[htbp]
\begin{center}
\FigureFile(14cm,14cm){thickTargetEmission2.eps}
\caption{The same as figure \ref{fig.BremsMonoEnergySpectrum}, 
but obtained under the thick-target condition 
using the thick atmosphere model. 
The solid curves represents analytic calculations 
by equation (\ref{eq.MonoBremsThick}). 
}
\label{fig.MonoThickSim}
\end{center}
\end{figure}

\clearpage
\begin{figure}[htbp]
\begin{center}
\FigureFile(14cm,14cm){thikTargetPow2.eps}
\caption{The same as figure \ref{fig.powerThinSpectrum}, 
but obtained under the thick-target condition 
using the thick atmosphere model. 
The solid curves represents analytic calculations 
by equation (\ref{eq.PowBremsThick}). }
\label{fig.PowThickSim}
\end{center}
\end{figure}

\clearpage
\begin{figure}[htbp]
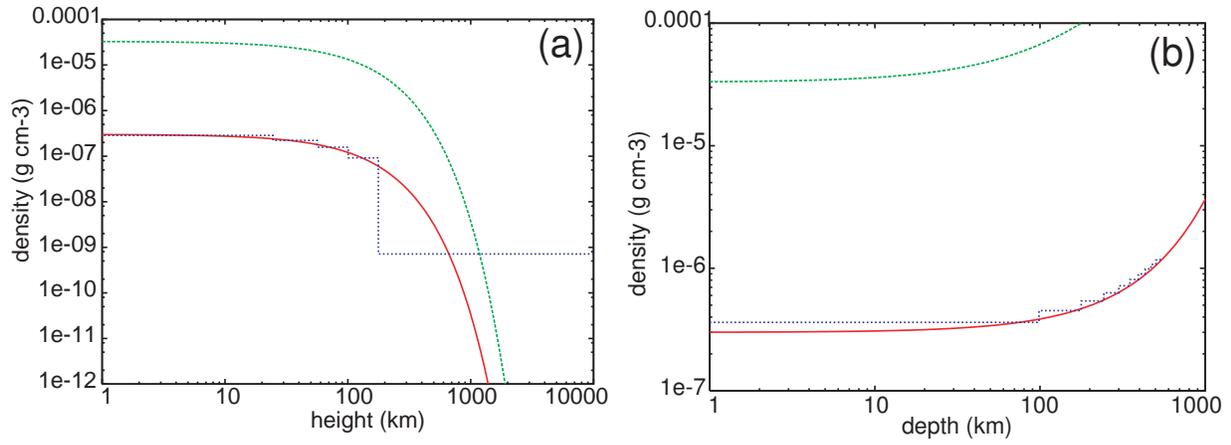

\begin{center}
\FigureFile(8cm,8cm){densityModel.eps}
\FigureFile(8cm,8cm){innerModel.eps}
\caption{Density profiles of our ``solar atmosphere'' (blue) 
compared with the analytical approximation by equation (\ref{eq.harvardDensity}), in the regions above the photosphere ($z>0$) panel (a) 
and below the photosphere ($z<0$) panel (b). 
Green curves show the column density integrated downwards from $z=+\infty$. 
}
\label{fig.zones}
\end{center}
\end{figure}

\clearpage
\begin{figure}[htbp]
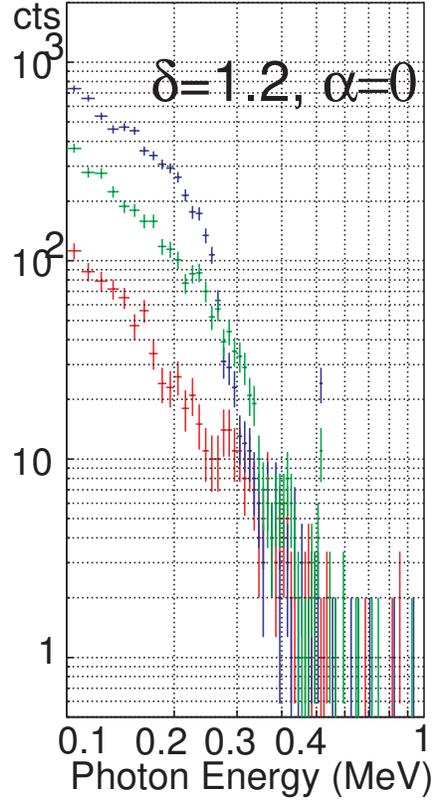

\begin{center}
\FigureFile(7cm,7cm){V9.241to100MeVG1.2e1000000rangePlist0Degat10000km_forPaper_mod.eps}
\caption{Monte-Carlo simulated spectra of gamma-rays emergent from our  ``solar atmosphere'', 
when 1-100 MeV electrons with $\delta=1.2$ are vertically injected into it. 
The photons, mostly Compton back-scattered, are averaged over the observer's azimuth $\varphi$ (figure \ref{fig.coordinate}), and collected at different viewing angle $\beta$; red, green, and blue crosses represent the photons satisfying $0.1< \cos\beta <0.2$, $0.5< \cos\beta <0.6$, and $0.9< \cos\beta <1.0$, respectively.  
}
\label{fig.ComptonGamma0Deg}
\end{center}
\end{figure}

\clearpage
\begin{figure}[htbp]
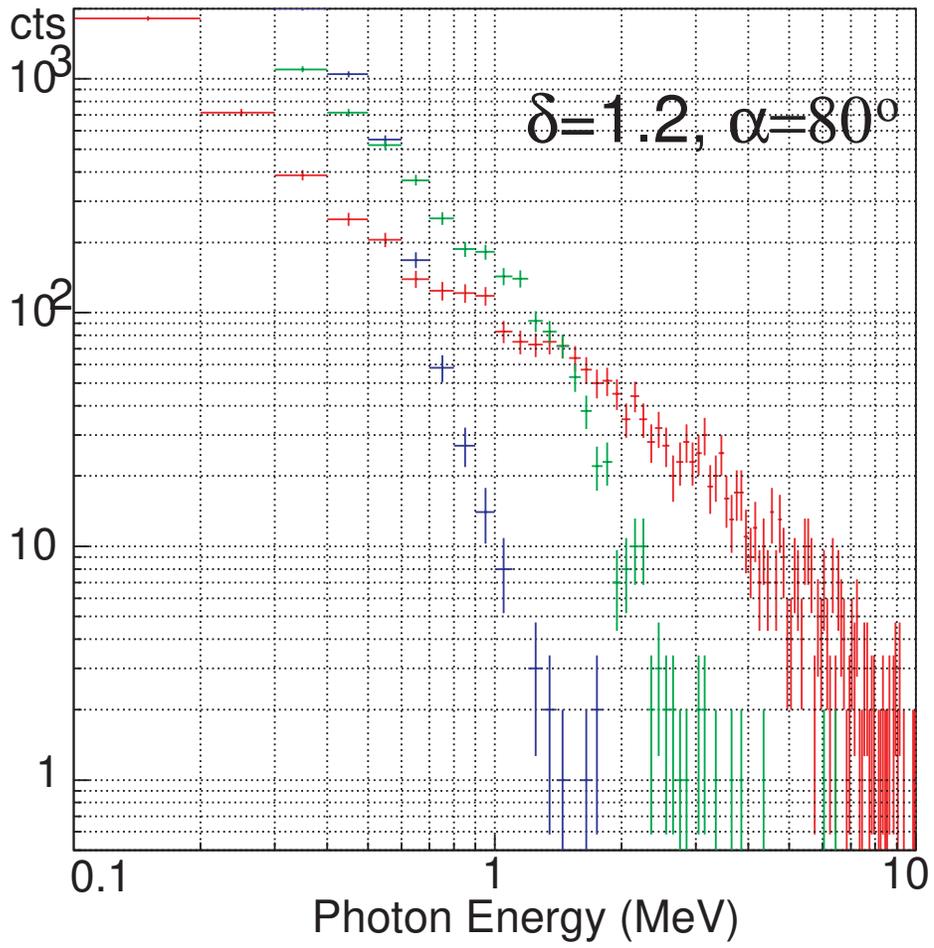

\begin{center}
\FigureFile(14cm,14cm){V9.241to100MeVG1.2e1000000rangePlist80Degat100km_forPaper.eps}
\caption{The same as figure \ref{fig.ComptonGamma0Deg},  
but simulating a highly slant injection with $\alpha = 80^{\circ}$ 
instead of the vertical injection. 
Note that the displayed energy range is different from that used in figure \ref{fig.ComptonGamma0Deg}
}
\label{fig.ComptonGamma80Deg}
\end{center}
\end{figure}

\clearpage
\begin{figure}[htbp]
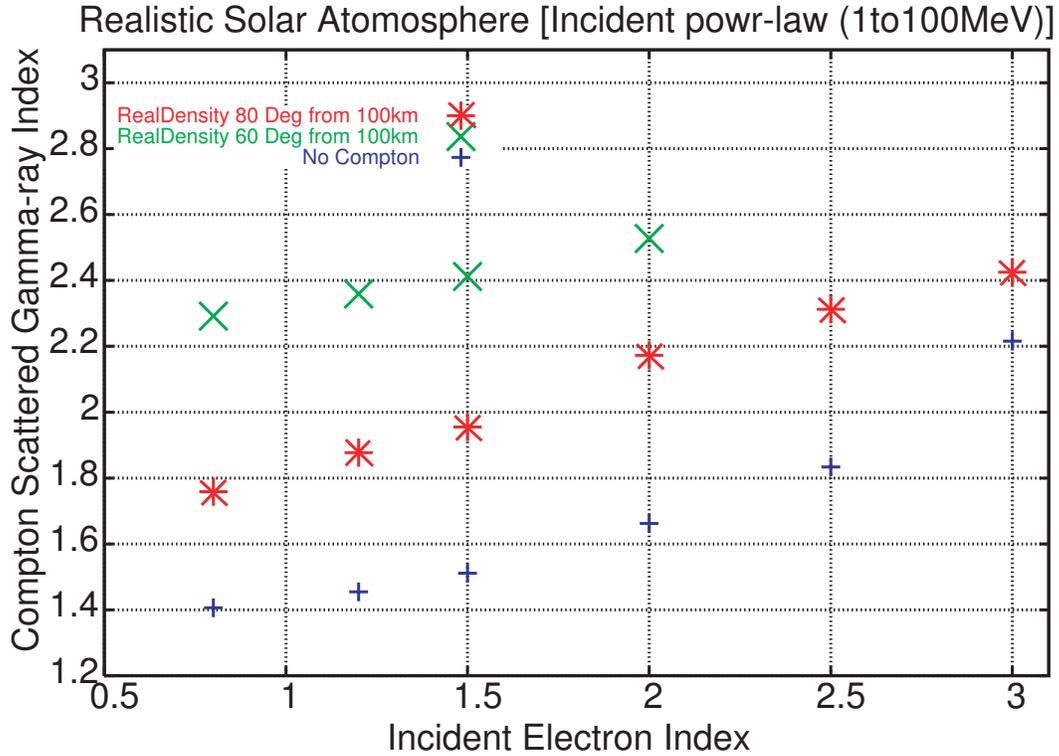

\begin{center}
\FigureFile(14cm,14cm){powObservedPhotonAll8.eps}
\caption{Photon indicies $\gamma$ of the 1-10 MeV gamma-rays, 
simulated under two favorable slant-injection conditions, 
shown as a function of the electron power-law index $\delta$ (over 1-100 MeV). 
Red asterisks and green crosses represent $\alpha=80^{\circ}$ and $\alpha=60^{\circ}$, 
respectively. 
In both cases, the emergent photons are accumulated over $0.1<\cos\beta<0.2$ 
and $0<\varphi<2\pi$. 
Blue crosses are the same as figure \ref{fig.thinthick}b, 
namely the spherically integrated bremsstrahlung photon index 
before subjected to the Compton process.  
}
\label{fig.DeltaGamma}
\end{center}
\end{figure}

\clearpage
\begin{figure}[htbp]
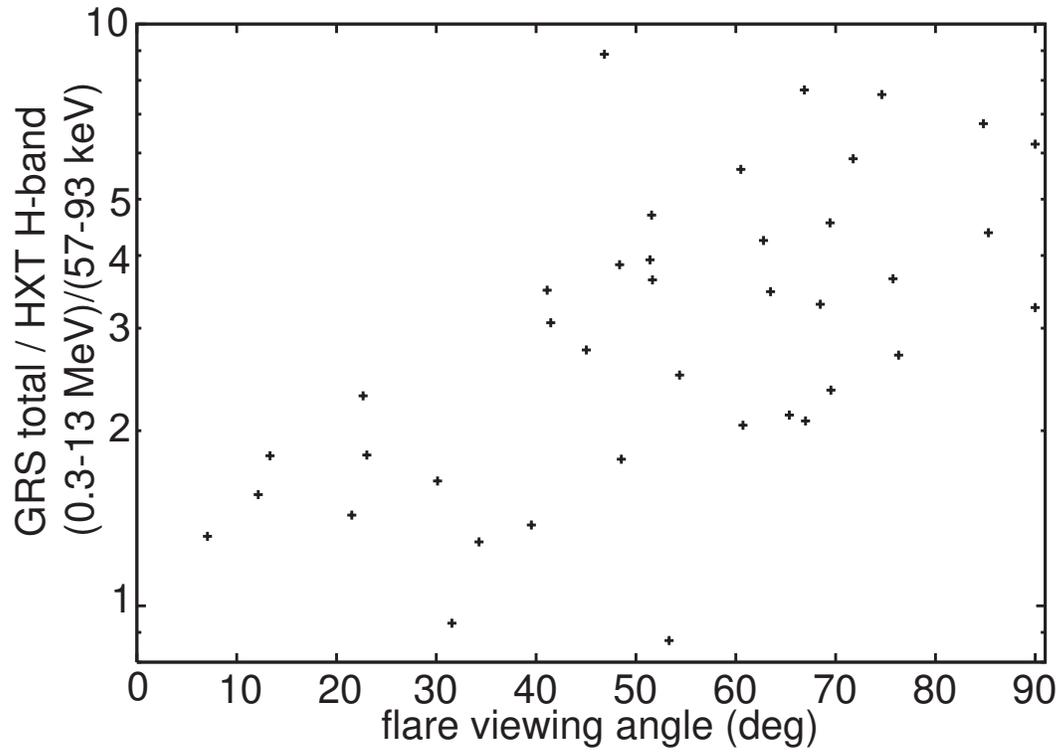

\begin{center}
\FigureFile(14cm,14cm){pos_vs_grs_per_hxMOD5.eps}
\caption{The gamma-ray (0.3--13 MeV) vs. hard X-ray (53--93 keV) peak count ratios of the 40 gamma-ray emitting solar flares observed by Yohkoh \citep{Yukari2005PASJ},
plotted as a function of their flare viewing angles calculated via equation (\ref{eq.theta_v}).}
\label{fig.longitude}
\end{center}
\end{figure}

\clearpage
\begin{figure}[htbp]
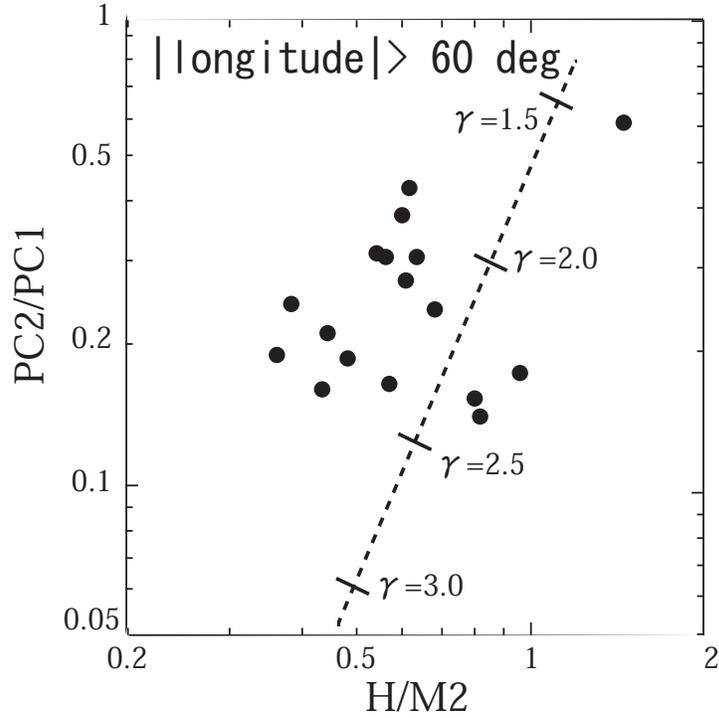

\begin{center}
\FigureFile(14cm,10cm){GRS_vs_HXT_limb.eps}
\caption{Color-color plots of the 40 gamma-ray emitting solar flares with absolute longitudes $> 60^{\circ}$, observed with Yohkoh. 
Abscissa is the hard X-ray color between the H-band (57---93 keV) 
and M2-band (33---57 keV) of the HXT, 
while ordinate is the gamma-ray color 
between the PC2-band (1.43---6.21 MeV) vs. PC1-band (0.22---1.43 MeV) of the GRS. 
The dashed line indicates the locus on which the two colors are represented 
by a single photon index $\gamma$. 
}
\label{fig.color_color2}
\end{center}
\end{figure}

\clearpage
\begin{figure}[htbp]
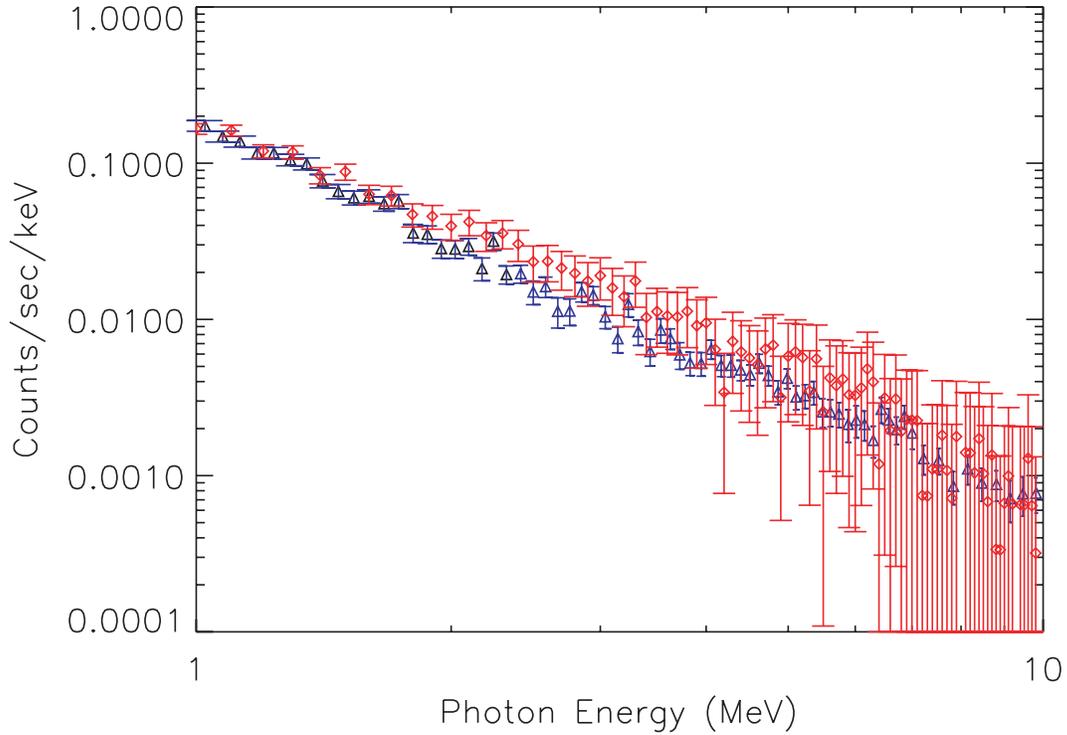

\begin{center}
\FigureFile(14cm,10cm){grs_sim_comp_D2.0_b2.eps}
\caption{A gamma-ray spectrum (blue) of the 1998 November 22 solar flare,
acquired with the Yohkoh GRS over a time period of  06:37--06:40 UT. 
It is compared with a simulation (red crosses),
computed assuming that $10^7$ electrons with a spectrum of $\delta=2.0$ (1--100 MeV) 
and  an incident angle distribution of 
equation~(\ref{eq.DermerRamaty1986}) are injected,
and the emergent photons are observed with $0.1 \le \cos \beta \le 0.2$.
The simulated spectrum has been convolved with an approximate
response of the GRS.
The normalization of the simulated spectrum is arbitrarily adjusted to match  the observed data.}
\label{fig.grs_sim_comp}
\end{center}
\end{figure}

\end{document}